\documentclass[groupedaddress,showpacs,aps,preprint,prb,floatfix]{revtex4-2}
\usepackage{bm,amsmath,amssymb}
\usepackage{graphicx}
\usepackage{xcolor}
\usepackage{textgreek}

\begin{document}

\title{Josephson-like tunnel resonance and large Coulomb drag in GaAs-based electron-hole bilayers}

\author{M. L. Davis}
\author{S. Parolo}
\author{C. Reichl}
\affiliation{Solid State Physics Laboratory, ETH Z\"urich, CH-8093 Z\"urich, Switzerland}
\author{W. Dietsche}
\affiliation{Solid State Physics Laboratory, ETH Z\"urich, CH-8093 Z\"urich, Switzerland}
\affiliation{Max-Planck-Institut f\"ur Festk\"orperforschung, D-70569 Stuttgart, Germany}
\author{W. Wegscheider}
\affiliation{Solid State Physics Laboratory, ETH Z\"urich, CH-8093 Z\"urich, Switzerland}
\affiliation{Quantum Center, ETH Z\"urich, CH-8093 Z\"urich, Switzerland}

\begin{abstract}

Bilayers consisting of two-dimensional (2D) electron and hole gases separated by a 10 nm thick AlGaAs barrier are formed by charge accumulation in epitaxially grown GaAs. Both vertical and lateral electric transport are measured in the millikelvin temperature range. The conductivity between the layers shows a sharp tunnel resonance at a density of $1.1 \cdot 10^{10} \text{ cm}^{-2}$, which is consistent with a Josephson-like enhanced tunnel conductance. The tunnel resonance disappears with increasing densities and the two 2D charge gases start to show 2D-Fermi-gas behavior. Interlayer interactions persist causing a positive drag voltage that is very large at small densities. The transition from the Josephson-like tunnel resonance to the Fermi-gas behavior is interpreted as a phase transition from an exciton gas in the Bose-Einstein-condensate state to a degenerate electron-hole Fermi gas.

\end{abstract}

\maketitle

The search for an excitonic Bose-Einstein condensate (BEC) \cite{BLATT1962} in electron-hole bilayers (EHBs) formed by closely spaced two-dimensional electron and hole gases (2DEGs and 2DHGs, respectively) has been ongoing for some decades \cite{LOZOVIK1975,Shev1976}. It is theoretically predicted to be a superfluid state whose signatures would not only include the dissipationless flow of the excitons but also a Josephson-like enhanced tunneling between the layers and a large Coulomb drag between the layers that persists down to the lowest temperatures \cite{Vignale1996,Littlewood_1996,Chang_2013,Perali2013,Fogler2014,Skinner2016,Su2017,Efimkin2020,Zeng2020,Saberi2020}.

The predicted BEC has been the objective of many publications dealing with a large variety of material systems. Both optical excitation and suitable doping in heterostructures have been used to produce excitons \cite{SIVAN1992,Snoke2002,Butov2002,Deng02,Seamons2009,Croxall2008}. Only very recently, some promising results have been reported in two-dimensional atomic layers \cite{Burg2018,Wang2019,Ma2021}. Additionally, evidence for a condensation has been found in 2DEG bilayers in quantizing magnetic fields \cite{Spi2000,Tut2004,Hyart2011}, which is of the Bardeen-Cooper-Schrieffer (BCS) type \cite{Ding2016}.

In this letter we use GaAs-based devices in which the two components of the EHB are separated by a $\text{Al}_{0.8}\text{Ga}_{0.2}\text{As}$ barrier and control the charge densities electrically without optical excitation. We use barriers of only 10 nm thickness, which is considerably thinner than what has been used in earlier experiments \cite{SIVAN1992,Pohlt2002,Croxall2008,Seamons2009}. The high quality of our heterostructures allows for the fine-tuning of the charge-carrier densities down to below 1$\cdot 10^{10} \text{ cm}^{-2}$. At these densities the ratio of Coulomb energy to kinetic energy, usually stated as the $r_{s}$ value, is sufficiently large to reach the Mott transition between an exciton gas and an electron-hole Fermi gas \cite{Fisher1988,Fogler2014,Saberi2020}. Near the transition we observe a large resonance in the tunnel current between the layers at zero magnetic field indicating a significantly increased tunneling between the electron and the hole layers, reminiscent of the Josephson tunneling observed in the excitonic condensate in 2DEG bilayers in quantizing magnetic fields \cite{Spi2000,Tut2004,Hyart2011}. There, the exciton condensation and the correlation between the two charge layers leads to a dramatic enhancement of the conductivity between the two layers. Our EHBs are different from those of the quantum Hall liquids because the electrons and holes are at different band-energies. A Josephson current would be a coherent recombination under the emission of a (coherent) photon as has been discussed by Xie et al. \cite{Xie2018} for an atomic bilayer system. We also observe an enhanced capacitance, or increased compressibility, in the same density range. This is expected for an exciton condensate as was pointed out by Ma et al. \cite{Ma2021}. At densities above the Mott transition the EHB transforms into 2D electron-hole Fermi gases that still show indications of Coulomb interaction like a large Coulomb drag between the layers. At even higher densities the mobility vs. density traces show novel interaction effects.

\begin{figure}
	\centering
	\includegraphics[width=0.95\columnwidth]{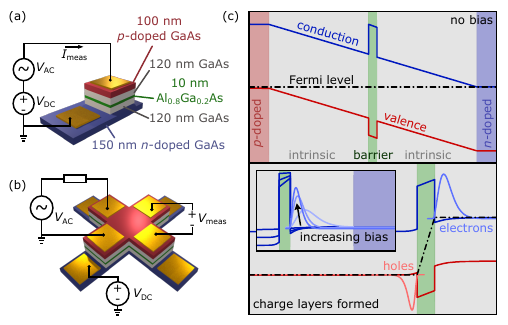}
	\caption{(a) Schematic of the pillar-shaped sample. The barrier (green) is sandwiched between undoped GaAs spacers. Holes and electrons are provided by \textit{p}- and \textit{n}-doped layers at the top and bottom, respectively. (b) Cross-shaped device with four contacts on each doped layer allowing lateral conductivity and frictional drag experiments to be performed on the EHB. (c) Band structure of the devices without a bias (top) and with a bias (bottom). After exceeding the flat-band condition charges accumulate at the barrier. The inset shows the simulated shape of the electron wavefunction at the barrier. The hole wavefunction has a similar shape.}
	\label{fig:LayersAndBandstructures}
\end{figure}

The heterostructures investigated in this study were grown by molecular-beam epitaxy. Two undoped GaAs layers separated by a 10 nm $ \text{Al}_{0.8}\text{Ga}_{0.2}\text{As}$ barrier are sandwiched between \textit{n}- and \textit{p}-bulk-doped GaAs layers, see Fig.~\ref{fig:LayersAndBandstructures}(a). Each 2D system, along with its respective doped layer, is contacted separately. Pillar devices with an area of $5.6 \cdot 10^{-4} \text{ cm}^{2}$ are used for capacitance and interlayer conductance measurements and have one contact on the top \textit{p}-layer, which does not penetrate the barrier. Cross-shaped samples (Fig.~\ref{fig:LayersAndBandstructures}(b)) have four contacts to both the \textit{n}-side and the \textit{p}-side, allowing for lateral transport and Coulomb drag measurements. The EHB area of the cross is 8.7$ \cdot 10^{-4} \text{ cm}^{2}$. Both devices were fabricated from the same growth run. For details of the technology see the supplement \cite{Sup}. Applying a DC voltage exceeding that corresponding to the GaAs band-gap energy leads to the flat-band condition and charges accumulate at the barrier, forming the EHB, see Fig.~\ref{fig:LayersAndBandstructures}(c). Adding a small AC voltage to the DC bias and measuring the in-phase and the out-of-phase components of the resulting AC current yields the tunnel conductance and the capacitance of the EHB, respectively \cite{Parolo2022}.

\begin{figure}
	\centering
	\includegraphics[width=0.95\columnwidth]{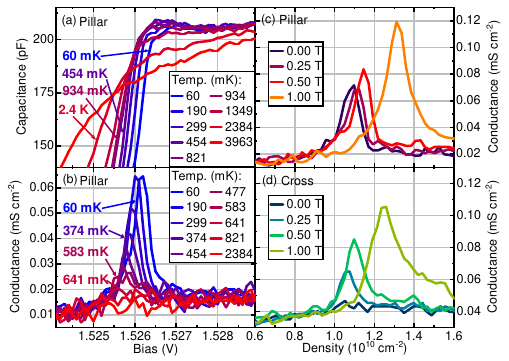}
	\caption{(a) Temperature dependence of the capacitance shoulder at the onset of mobile charge accumulation. (b) Tunnel resonance of the pillar sample. (c) Tunnel resonance with different magnetic fields applied (pillar sample). (d) Same as (c) but measured on the cross-shaped sample.}
	\label{fig:Josephson}
\end{figure}

Measurements of capacitance and conductance are done at the base temperature (20 mK) of a dilution refrigerator. This corresponds to a typical electron temperature of about 60 mK \cite{Sup}. Results measured at a frequency of 283 Hz and with an AC voltage amplitude of 50 \textmu V are shown in Fig.~\ref{fig:Josephson}. The capacitance increases step-like at about 1.525 V to values that correspond to a charge carrier separation of tens of nanometers and signal the existence of an EHB. Beyond the step, the capacitance continues to increase because the shifting of the wave functions leads to a decreasing effective interlayer distance, see Fig.~\ref{fig:LayersAndBandstructures}(c). The shape of the capacitance vs. bias curve does not depend on frequency between 30 Hz and about 3 kHz, indicating that the doped layers and 2D charge gases are in equilibrium in this frequency range. Around the capacitance step, the ordinary tunnel conductance across the barrier is 6 nS, corresponding to a conductivity of 10 \textmu $\text{S cm}^{-2}$. It increases exponentially with increasing bias as expected for Fowler-Nordheim tunneling across a narrow barrier. Resonant tunneling is not expected until much higher biases \cite{Finley1998}.

Both the capacitance and the tunnel conductance show unusual behavior in this bias region. A shoulder is observed in the capacitance while the tunnel conductance shows a striking resonance. Compared to the ordinary tunnel current across the barrier, the resonance peak is five times higher for our predominantly used measurement frequency of 283 Hz, but varies with frequency (see supplement \cite{Sup}).

In Fig.~\ref{fig:Josephson}(a) and (b) we plot the temperature dependence of the capacitance and the tunnel resonance, respectively. The capacitance shoulder degrades with increasing temperature and disappears above 2 K, which corresponds to a 100 \textmu eV energy scale, much smaller than typical energies in semiconductors like band offsets and the ionization energies of impurities. The tunnel resonance begins to diminish at a temperature of around 200 mK and continues to do so until it disappears entirely around 1 K. Thus the tunnel resonance exists only in roughly the same temperature range as the capacitance shoulder.

A capacitance enhancement has recently been interpreted as a sign of the increased compressibility that should accompany a BEC state in an EHB \cite{Ma2021}. Its occurrence in our samples supports our interpretation that the substantially enhanced conductance across an insulating barrier is a hallmark for Josephson-like charge exchange within an EHB in a condensed state \cite{Spi2000,Tie2008,Burg2018}.

\begin{figure}
	\centering
	\includegraphics[width=0.9\columnwidth]{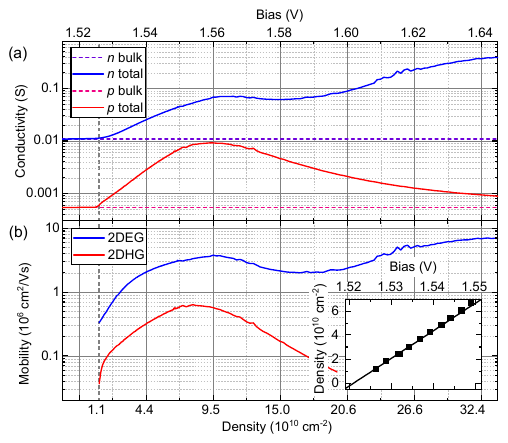}
	\caption{(a) Conductivities measured on the \textit{n}- and \textit{p}-doped sides of the device vs. bias voltage. (b) 2DEG and 2DHG mobilities as a function of density. (Inset) Density vs. bias voltage. The dots mark densities obtained from quantum-Hall-effect minima.}
	\label{fig:LateralConductance}
\end{figure}

Discussing the capacitance and tunneling behavior in the context of condensed excitons requires knowledge of the actual charge densities. The densities of the two 2D charge layers are separately accessible from the Shubnikov-de-Haas oscillations of their respective resistances (see the supplement \cite{Sup}). The oscillations in the \textit{n}- and \textit{p}-doped sides agree with each other, indicating that both 2D charge layers have the same density, as assumed. The densities at biases around $\text{1.53 V}$ are particularly relevant, since this is where the tunnel resonance of Fig.~\ref{fig:Josephson} is observed. This density range is accessible from capacitance vs. bias plots in moderate magnetic fields, which show minima at integer fillings of the Landau levels. Thus, the density as a function of bias is precisely determined down to 1.2 $\cdot 10^{10} \text{ cm}^{-2}  $. Smaller densities are estimated by extrapolation. A plot of density vs. bias at small densities is shown in the inset in Fig.~\ref{fig:LateralConductance}.

In Fig.~\ref{fig:Josephson}(c), the tunnel resonance from Fig.~\ref{fig:Josephson}(b) is plotted as a function of density. The density at the maximum of the resonance is 1.1$ \cdot 10^{10} \text{ cm}^{-2} $ with no magnetic field applied. Also plotted in Fig.~\ref{fig:Josephson}(c) is the effect of different magnetic fields oriented perpendicular to the EHB plane. The resonance grows and shifts slightly to higher densities with increasing field.

Fogler et al. \cite{Fogler2014} show a zero-temperature phase diagram of an EHB. At small densities (large $r_{s}$ values) a condensed exciton gas is expected. At larger densities the excitons are expected to separate and form electron and hole Fermi liquids. Under the conditions of our experiments $ r_{s} \approx 8$ and the ratio $d/a_{e} \approx 6$, which puts our system very near to the expected phase transition between an exciton gas and an electron-hole Fermi liquid. Here $r_{s} = (\pi n_{e}{a_{e}^{2}})^{-1/2}$, where $a_{e} \approx 7$ nm is the exciton Bohr radius, $ n_{e} $ is the electron-hole pair density, and $ d$ is the effective interlayer distance, which is extracted from bandstructure simulations and is 44 nm at the density of the tunnel resonance \cite{Nextnano2007}.

The tunnel resonance is also observed in the cross sample from Fig.~\ref{fig:LayersAndBandstructures}(b), at the same density of 1.1$ \cdot 10^{10} \text{ cm}^{-2} $ (see Fig.~\ref{fig:Josephson}(d)). This sample is intrinsically less homogeneous because, in contrast to the pillar, the active area is only partially covered by contact material. This is probably the reason why the tunnel resonance is very weak in zero magnetic field. It nevertheless reaches similar values as in the pillar device already at very moderate magnetic fields (0.25 T).

We identify the tunnel resonance and the capacitance shoulder with a Josephson-like coupling between the electrons and holes in an exciton condensate \cite{Burg2018,Xie2018}. The length scales are favorable for an exciton gas and a BEC state is expected in the low-temperature limit. Since a BEC requires a sufficiently high density to form at non-zero temperature, one may not be able to see the tunnel resonance at even lower densities. At higher densities a BEC is not possible because of the transition into an electron-hole Fermi gas where the charges behave as a 2DEG adjacent to a 2DHG.

The conductances of the 2D charge gases can be measured in the cross sample. The two doped layers form two crosses superimposed on top of each other. Separate contacts to the ends of the cross arms allow measurement of the conductance per square (or conductivity) of the two crosses separately by the van-der-Pauw method. Results are plotted as a function of bias in Fig.~\ref{fig:LateralConductance}(a). Clearly, the conductivities of both the \textit{n}-side and the \textit{p}-side increase at about the bias at which the step-like increase of the capacitance is observed. Below this threshold the respective conductivities are almost constant. It is straightforward to subtract the doped layer conductivities and find those of the 2D charge gases. Transport mobilities of both the 2DEG and the 2DHG are plotted in Fig.~\ref{fig:LateralConductance}(b).

 Both 2D gases show striking mobility maxima that are not observed in standard heterostructures \cite{Watson2011,Kuelah2021} nor were they reported in an earlier EHB publication \cite{Morath2008}. Mobility peaks of both the 2DEG and the 2DHG occur at a density ($ 9\cdot10^{10} \text{ cm}^{-2} $) at which the average interlayer distance, around 27 nm, is similar to the intralayer distance ($ \approx$ 33 nm) in both layers. Whether or not this is a coincidence will be clarified by future samples with different barrier thicknesses. Although the present devices are not optimized for high mobilities, they  reach rather high values - about 7 million$ \text{ cm}^{2}\text{ V}^{-1} \text{ s}^{-1} $ at a density of about $3.5\cdot 10^{11} \text{ cm}^{-2} $ for the 2DEG and 0.6 million$ \text{ cm}^{2}\text{ V}^{-1}\text{ s}^{-1} $ at a density of about  $9\cdot10^{10} \text{ cm}^{-2} $ for the 2DHG.

In the next part of this letter, we turn to measurements of frictional, or `Coulomb', drag between the two 2D charge layers. A  current $ I_{\text{drive}} $ flowing along one 2D charge gas of a bilayer system leads to a drag voltage $ V_{\text{drag}} $ in the second layer due to Coulomb interactions. This is described by the drag resistivity $\rho_{D} = (W/L) \cdot (V_{\text{drag}}/I_{\text{drive}} $), with $W$ and $L$ being the length and width of the EHB layers, respectively. In a Fermi gas, the drag effect originates from momentum transfer via scattering processes of the charge carriers, leading to a $\rho_{D} \propto T^{2} $ dependence, see e.g., \cite{Jauho1993,Hwang2018}. The magnitude of the drag can be a signature of a BEC state as has been demonstrated for coupled 2DEGs at half-filled Landau levels \cite{Eisenstein2014}.

\begin{figure}
	\centering
	\includegraphics[width=0.9\columnwidth]{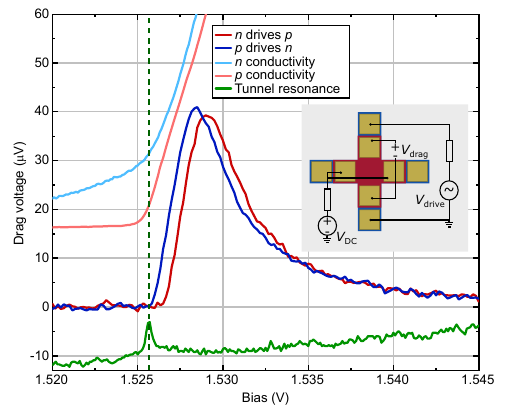}
	\caption{Drag voltages measured for the \textit{n}-side driving the \textit{p}-side and vice versa. The driving AC current in both cases is 60 nA. Also shown are the tunnel resonance and the conductivities of the \textit{n}- and \textit{p}-layer of the same sample (both in arbitrary units).}
	\label{fig:FrictionalDrag}
\end{figure}

We use the cross sample to measure the frictional drag between the 2DEG and the 2DHG. An AC current of 60 nA and 13.8 Hz passes straight across one layer while the drag voltage is picked up in the parallel layer, see inset of Fig.~\ref{fig:FrictionalDrag}. A large resistor (100 k$\Omega $) is used in series with the DC voltage source to reduce any AC-current coupling between the two layers via the voltage source \cite{DasGupta2011,NPRHill1996}. Using a significantly smaller or no resistor at all was found to influence the magnitude of the measured drag voltage.

Drag voltages as a function of bias are plotted in Fig.~\ref{fig:FrictionalDrag}. The data show that a positive drag voltage is indeed observed as soon as the 2D gases show an increased conductance as signaled by kinks in the \textit{n}- and \textit{p}-layer conductivities. Inverting the drive and the drag layers leads to qualitatively and quantitatively similar curves, in contrast to earlier drag data where significant asymmetries have been observed \cite{Seamons2009,Croxall2008}. The signal of the dragged \textit{p}-layer is shifted to slightly higher biases, probably due to the intrinsically lower conductivity of the 2D hole layer compared to the 2DEG. The drag voltages reach maxima at ca. 1.529 V and subsequently decrease with increasing bias. The sign of the drag voltages is positive, which is expected for different polarities in the drag and drive layers. It is significant that the DC bias (and with that the 2D density) at which the steep rise in lateral conductivity occurs coincides not only with the onset of the positive drag voltage but also with the tunnel resonance.

We rule out that the positive drag voltages are caused by the interlayer conductance. An estimate of the expected spurious voltage produced by a network of the \textit{p}- and \textit{n}-layers and the tunnel resistance in the relevant bias regime leads to drag resistances of the order of $ 10^{-3} \text{ }\Omega $.

\begin{figure}
	\centering
	\includegraphics[width=0.9\columnwidth]{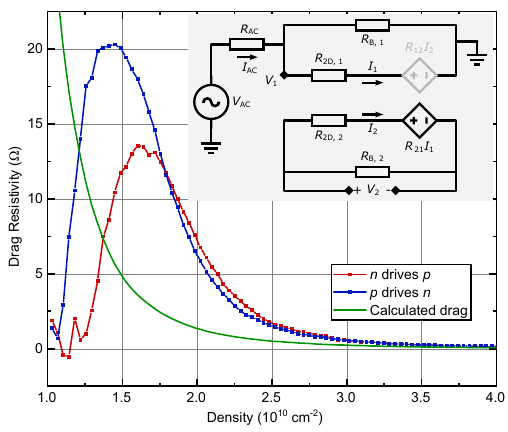}
	\caption{Intrinisic drag resistivities for the 2DHG driving the 2DEG and vice versa. The green line shows the expected Coulomb drag between two Fermi gases according to the theoretical model. The inset shows the circuit used to extract the intrinsic drag resistivities from the measured drag voltages.}
	\label{fig:RealDrag}
\end{figure}

The measured drag voltages do not correspond directly to the intrinsic drag resistivities between the 2D charge gases because the conducting doped layers modify both the actual current flowing along the drive 2D layer and the measured voltage on the drag side. It follows from a linear network model, see inset of Fig.~\ref{fig:RealDrag}, that the intrinsic drag resistivity ($ \varrho_{D} $) is related to the measured drag resistance ($ R_{D} $) by the relation 
$ \varrho_{D}=R_{D}\cdot \frac{W}{L} \cdot (R_{\text{2D,1}}+R_{\text{B,1}})\cdot (R_{\text{2D,2}}+R_{\text{B,2}})/(R_{\text{B,1}}\cdot R_{\text{B,2}}) $,
where $ R_{\text{2D,1}} $ and $ R_{\text{2D,2}} $ are the resistances of the 2D drive and drag layers in the EHB, respectively, and $R_{\text{B,1}}  $ and $ R_{\text{B,2}} $ those of the two doped layers, see the supplement for details \cite{Sup}.

In Fig.~\ref{fig:RealDrag} the intrinsic drag resistivities are plotted as a function of density, reaching maxima at 13 $ \Omega $ and 20 $ \Omega $. This difference is a result of the correction procedure detailed above; it is not clear if this is intrinsic to the device or if the correction formalism needs to be improved. The solid green line corresponds to the density-dependent Coulomb drag resistivity between two dilute 2D systems expected from theory at 60 mK and adjusted for the interlayer distances in our experimental setup \cite{Hwang2003}. It is in qualitative agreement with our experimental data for high densities above $ 4\cdot 10^{10} \text{ cm}^{-2} $. Below that, we observe a significantly enhanced drag resistivity as the density is reduced and the 2D layers approach the Mott transition into the excitonic state. The signal peaks at a density of ca. $ 1.5\cdot 10^{10} \text{ cm}^{-2} $ and is subsequently cut oﬀ due to the very small individual conductivities of the two 2D layers that make the drag experiment impossible.

In conclusion, we prepared high-mobility electron-hole bilayers with barriers of only 10 nm thickness. In these structures we can precisely control densities down to $ 1\cdot 10^{10} \text{ cm}^{-2} $, thus allowing access to the excitonic and the Fermi gas regime of the bilayer system on the low- and high-density side of the Mott transition, respectively. We measure vertical and lateral transport and find, near the Mott transition, a capacitance enhancement and a strongly increased conductance that we interpret as the enhanced compressibility and the Josephson-like charge transfer in an exciton condensate, respectively. In future experiments, increasing the barrier thickness and applying an in-plane magnetic field will both be used to test their effect on a Josephson coupling. Measuring the counterflow conductance in the excitonic regime could demonstrate excitonic flow that is not yet accessible.

\begin{acknowledgements}
We thank Klaus von Klitzing for illuminating discussions and Peter Märki for sharing his technical expertise and providing valuable support. We acknowledge financial support from the Swiss National Science Foundation (SNSF) and the National Center of Competence in Science ``QSIT - Quantum Science and Technology".

M.L.D., S.P., and C.R. performed measurements and analysis. M.L.D. and S.P. fabricated the samples and C.R. grew the wafer. W.D. and W.W. conceived of the experiment and interpreted the results. W.D. wrote the manuscript, M.L.D. wrote the supplement, and C.R. made contributions to both.
\end{acknowledgements}

\end{document}


\title{Supplemental material: Josephson-like tunnel resonance and large Coulomb drag in GaAs-based electron-hole bilayers}

\author{M. L. Davis}
\author{S. Parolo}
\author{C. Reichl}
\affiliation{Solid State Physics Laboratory, ETH Z\"urich, CH-8093 Z\"urich, Switzerland}
\author{W. Dietsche}
\affiliation{Solid State Physics Laboratory, ETH Z\"urich, CH-8093 Z\"urich, Switzerland}
\affiliation{Max-Planck-Institut f\"ur Festk\"orperforschung, D-70569 Stuttgart, Germany}
\author{W. Wegscheider}
\affiliation{Solid State Physics Laboratory, ETH Z\"urich, CH-8093 Z\"urich, Switzerland}
\affiliation{Quantum Center, ETH Z\"urich, CH-8093 Z\"urich, Switzerland}

\maketitle

\section{Devices}

The GaAs/AlGaAs heterostructures investigated in this work were grown by molecular beam epitaxy in a modified Varian Gen II setup optimized for highest material quality, capable of producing two-dimensional-electron-system (2DES) structures with mobilities exceeding $30\cdot10^6\text{ cm}^2/\text{Vs}$ \cite{Kuelah2021}. GaAs substrates were overgrown at a temperature of 625$^{\circ}$C under continuous rotation. Underneath the electron-hole-bilayer (EHB) region, a 1000 nm thick superlattice of alternating $\text{Al}_{0.3}\text{Ga}_{0.7}\text{As}$ and GaAs was grown. This sequence is necessary to reduce the initial surface roughness and impurity levels of the commercially available substrates. Following the buffer stack, 150 nm of \textit{n}-type Si:GaAs form the backgate. The doping density is gradually lowered in several steps towards the EHB to reduce the total amount of dopants needed, for an average doping density of $2.1\cdot10^{18}\text{ cm}^{-3}$. This was done in an attempt to minimize the influence of the bulk doping on the measurements of the 2D charge carrier systems. The EHB region itself consists of two layers of 120 nm undoped GaAs each, separated by a tunnel barrier of seven thin AlAs layers (1.132 nm), with one monolayer (0.283 nm) of GaAs in between. The total thickness of the barrier is 9.6 nm, with an average aluminum content of 82\%. Above the EHB another bulk-doped layer, 100 nm of \textit{p}-type C:GaAs, was placed, likewise with gradually decreasing density towards the barrier for an average of $4.3\cdot10^{18}\text{ cm}^{-3}$. Placing the carbon doping in GaAs instead of AlGaAs sets this heterostructure apart from those used in an earlier study, where the 2DHG was located in a rectangular quantum well, which led to a significantly different behavior of the EHB \cite{Parolo2022}.\par

The grown heterostructure was processed into two different device geometries – pillar and cross – using standard optical lithography and chemical wet etching. For the pillar device, a mesa of 285 x 195 \textmu m$^2$ was covered with photoresist and the surrounding semiconductor stack etched to a depth of about 300 nm, into the undoped GaAs below the tunnel barrier. Then a single \textit{p}-type Ohmic contact stack was evaporated on top of the mesa, and an \textit{n}-type contact onto the etched surface nearby. For the second device, a 450 nm deep etch was performed around a symmetric cross shape with a diameter of 670 \textmu m and arm widths of 130 \textmu m. Then, a second cross with a diameter of 400 \textmu m and arm widths of 170 \textmu m was placed centered on the first one, and the area around it etched down to roughly 290 nm, well below the tunnel barrier. This etch step defines the area of the EHB. Ohmic contacts were subsequently evaporated near the outer edges of the cross arms, \textit{p}-type on the top, \textit{n}-type on the remaining areas of the lower cross. The \textit{n}-type Ohmic stack consists of Au/Ge/Ni, thermally annealed for 300 s at 475$^{\circ}$C. The \textit{p}-type contacts are composed of Ti/Pt/Au, annealed for only 30 s at 300$^{\circ}$C. This combination of layer stack, annealing time, and temperature ensured a reliable electrical contact to the \textit{p}-side of the device, while the contact remained isolating to the \textit{n}-side.\par

\section{Measurement Setup}
\label{sec:MeasurementSetup}

All measurements presented in this work were performed in a Triton200 dilution refrigerator with a base temperature of 20 mK. However, due to practical limitations in the thermalization of the measurement lines, the lowest achievable temperature of the charge carrier systems investigated is significantly higher, around 60 mK. This value has been determined by comparing the magnetotransport characteristics of a high-mobility 2DES against a set of measurements performed on the same device, but in a different refrigerator with carefully calibrated electron temperatures. Specifically, the temperature dependence of the fractional quantum Hall effect features at and near filling factor \textnu\space = 5/2 were used. The 5/2 state, as well as the four adjacent reentrant integer quantum Hall states are known to be extremely fragile \cite{Reichl_2014,PhysRevLett.108.086803,PhysRevB.81.035316} and thus are sensitive to small changes of the electron temperature even in the low millikelvin range.\par

\section{Measurement Circuits}

\subsection{Interlayer Conductance and Capacitance}
\label{sec:InterlayerCircuit}

\begin{figure}
	\centering
	\includegraphics[width=0.9\columnwidth]{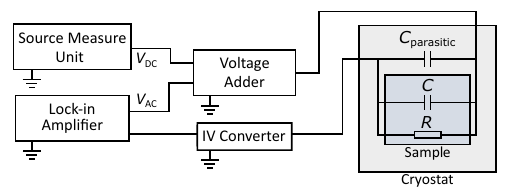}
	\caption{Circuit used for interlayer conductance and capacitance measurements. The sample is assumed to behave as a capacitor and a resistor in parallel.}
	\label{fig:SM_InterlayerMeasDiagram}
\end{figure}

Fig.~\ref{fig:SM_InterlayerMeasDiagram} shows the circuit used for interlayer measurements. A voltage is applied over the tunnel barrier of the device (one terminal connected to the \textit{p}-side contacts, the other to the \textit{n}-side contacts) and the resulting current measured. The voltage is comprised of a relatively small AC excitation voltage, provided by a lock-in amplifier, added to a DC bias voltage, provided by a source measure unit (SMU). The current through the device is measured using the lock-in supplying the AC excitation voltage, yielding differential in-phase and out-of-phase components. We assume the device behaves approximately as a resistor ($R$) in parallel with a capacitor ($C$), such that its capacitance and differential conductance can be extracted from this measurement. For a discussion of this assumption, see section \ref{sec:Frequency}. Note that the unshielded measurement lines add a parasitic capacitance, which is parallel to the device capacitance and thus produces a constant offset in the total measured capacitance.\par

\subsection{Intralayer Resistivity}
\label{sec:IntralayerCircuit}

\begin{figure}
	\centering
	\includegraphics[width=0.9\columnwidth]{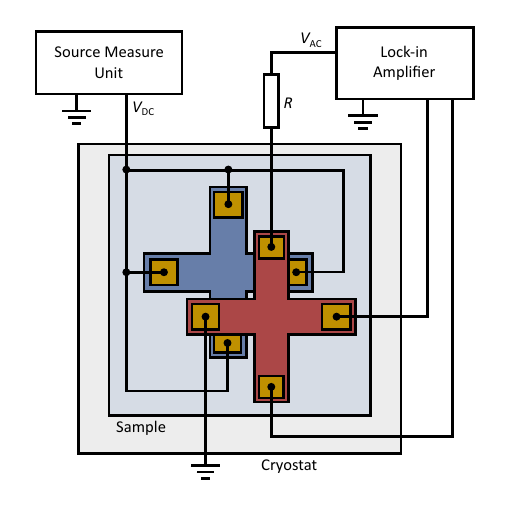}
	\caption{Circuit used for intralayer resistivity measurements. The \textit{n}-side (in blue) and \textit{p}-side (in red) of the sample are represented as two offset crosses.}
	\label{fig:SM_IntralayerMeasDiagram}
\end{figure}

Intralayer measurements were performed to determine the resistivity of the individual layers. In Fig.~\ref{fig:SM_IntralayerMeasDiagram} the circuit for measuring the \textit{p}-side resistivity is shown. The DC voltage bias is applied to the \textit{n}-side contacts through the SMU. An AC excitation voltage from the lock-in is applied between two neighboring \textit{p}-contacts, over a preresistor large enough to define the measurement current. The voltage is then measured across the remaining contacts of the \textit{p}-side. This is done for all (four) permutations to collect a full set of four-point resistances, from which the resistivity can be calculated using the van-der-Pauw formula. The \textit{n}-side resistivity is measured analogously.\par

\section{Density and Mobility Determination}
\label{sec:DensityAndMobility}

\begin{figure}
	\centering
	\includegraphics[width=0.9\columnwidth]{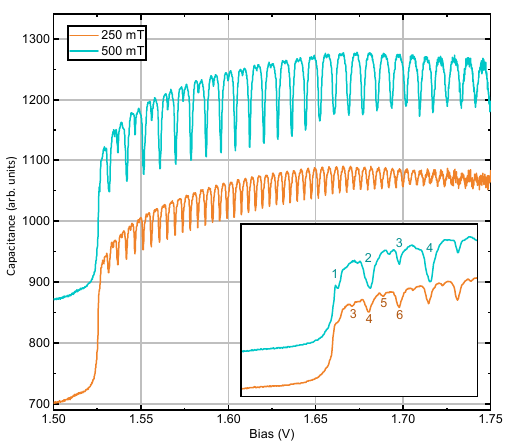}
	\caption{Oscillations in the capacitance at 250 mT and 500 mT applied magnetic field values (cross sample). The 500 mT curve is offset for clarity. Filling factors are assigned to the dips in order to determine the density at different biases. The inset shows the filling factors assigned to the dips at low bias.}
	\label{fig:SM_DensityFromSdH}
\end{figure}

In order to determine charge carrier densities in the 2D layers, interlayer capacitance measurements were performed in the presence of a magnetic field. Under these conditions, tuning the 2D-layer densities by sweeping the DC bias leads to quantum capacitance oscillations, which correspond to the integer Landau level fillings, and are equivalent to Shubnikov-de-Haas oscillations. The charge carrier densities at the bias values of these oscillations can thus be precisely determined. This was done for the relatively small fields of 250 mT and 500 mT, in order to get a large number of densities. Fig.~\ref{fig:SM_DensityFromSdH} displays these measurements, with the inset showing some examples of the filling factor assignments.\par

\begin{figure}
	\centering
	\includegraphics[width=0.9\columnwidth]{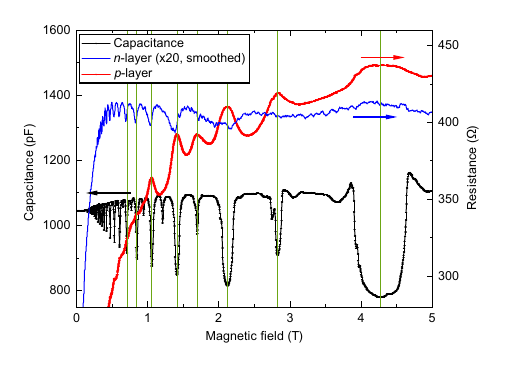}
	\caption{Oscillations of the interlayer capacitance and of the \textit{p}- and \textit{n}-layer resistances coincide, showing that the 2DEG and 2DHG densities are identical. The density here is $\text{20.6} \cdot \text{10}^{\text{10}} \text{ cm}^{-2} $ at a bias of 1.6 V.}
	\label{fig:SM_EqualDensity}
\end{figure}

We found the densities to increase approximately linearly with bias (see Fig. 3 of the main text). As is reflected in the capacitance vs. bias measurements (see Fig.~\ref{fig:SM_DensityFromSdH}), the slope of this linear relationship does increase slightly with bias. Also, our method has a minimum density (at a minimum bias) that can be directly calculated. For even lower biases/densities, we use a linear extrapolation based on the ten lowest measured densities. This analysis has been performed for each sample individually. Note that the oscillations in the capacitance coincide with those of the \textit{p}- and \textit{n}-layer resistances (see Fig.~\ref{fig:SM_EqualDensity}), confirming that the densities of the 2DEG and 2DHG are equal. In high fields the parallel conductance of the doped layers leads to a moderate quality of the oscillations, which nevertheless remain clearly visible.\par

With the densities determined, the remaining inputs needed for mobility calculations are the resistivity values of the 2D layers. It is safe to assume that the resistivity of both bulk layers remains constant over the entire bias range, and is equal to the total resistivity measured at low biases (before the 2D layers have formed). We invert the measured resistivity to obtain the total conductivity, then extract the conductivity of the 2D electron layer by subtracting the constant conductivity of the \textit{n}-doped bulk layer from the total. In the same manner we obtain the conductivity of the 2D hole layer.\par

\begin{figure}
	\centering
	\includegraphics[width=0.9\columnwidth]{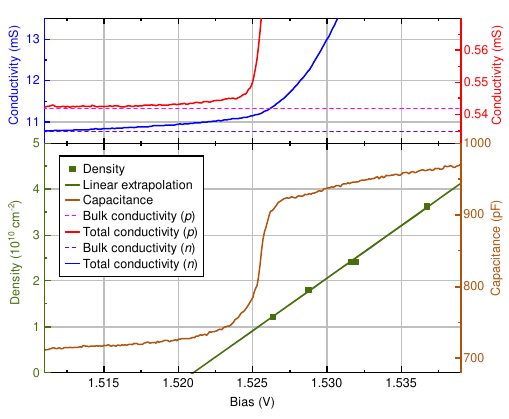}
	\caption{Capacitance and intralayer conductivity curves compared to the calculated density values (cross sample).}
	\label{fig:SM_IntermediateRegime}
\end{figure}

The \textit{p}-side and \textit{n}-side conductivities both experience an upturn at a DC bias of around 1.525 V. However, prior to this point they are already seen to rise with a relatively small slope (see Fig.~\ref{fig:SM_IntermediateRegime}). We attribute this to electrons and holes moving into localized states at the barrier at small biases, allowing for only limited conductivity. After around 1.525 V, true 2D charge gases form and the conductivity begins to increase substantially. The relatively modest rise in the capacitance could be due to a low density of these localized states (i.e. a low quantum capacitance). If this theory is correct, the linear extrapolation of the densities will be inaccurate for biases below around 1.525 V. In the main text (see Fig. 3) we have therefore presented the mobility only for larger biases.\par

\section{Frequency Dependence}
\label{sec:Frequency}

In our determination of the interlayer capacitance and differential conductance via scaling the measured current, we implicitly assume our device behaves as a capacitor in parallel with a resistor at our chosen frequency (283 Hz). To test this assumption, we also measured at significantly lower and higher frequencies (30 and 2833 Hz). The expectation would then be that the in-phase differential current remains constant, while the out-of-phase component scales linearly with the chosen excitation frequency. For the most part, this is what we observe; however, there are some deviations from this behavior, which we describe below.\par

\begin{figure}
	\centering
	\includegraphics[width=0.9\columnwidth]{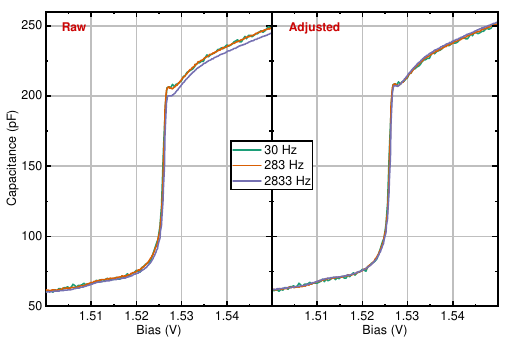}
	\caption{Frequency dependence of the calculated capacitance (pillar sample).}
	\label{fig:SM_FreqCapacitance}
\end{figure}

\begin{figure}
	\centering
	\includegraphics[width=0.9\columnwidth]{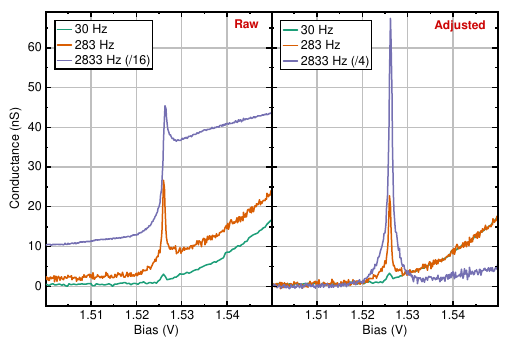}
	\caption{Frequency dependence of the calculated differential conductance (pillar sample). The 2833 Hz curves have been scaled down so that they can be easily viewed on the same plots as the other curves.}
	\label{fig:SM_FreqConductance}
\end{figure}

Fig.~\ref{fig:SM_FreqCapacitance} and Fig.~\ref{fig:SM_FreqConductance} show the capacitance and differential conductance, respectively, measured at all three frequencies. In the right panels, the curves have been adjusted to account for a phase shift of the signal caused by the measurement equipment. A first estimate of this phase shift is obtained by taking measurements with a test load. Further adjustments of the phase are then performed based on the assumption that at low bias the conductance should be close to zero. Taking $\sim$500 \textOmega \space as our series resistance from the measurement lines and $\sim$60 pF as the device capacitance leads to an effective parallel conductance lower than 1 nS at the highest frequency, justifying this assumption. Both the phase shift and the out-of-phase current are the largest for the measurements taken at 2833 Hz, resulting in the appearance of the capacitance signal in that of the conductance (compare Fig.~\ref{fig:SM_FreqCapacitance} and the left panel of Fig.~\ref{fig:SM_FreqConductance}).\par

After the adjustment the capacitance values for all three frequencies match, corresponding to the expected linear increase in out-of-phase signal with frequency. This demonstrates that the charges forming the EHB are in equilibrium with those of the doped layers from 30 Hz to nearly 3 kHz. We attribute this to the flat conduction and valence bands in the undoped GaAs, which prevent any slow relaxation processes. The conduction values, on the other hand, are not equivalent for all biases. We can distinguish two main parts of the conductance signal, one that increases continuously with bias (we call this the `background'), the other the tunnel peak. The adjustment results in an alignment of the conductance background. However, the conductance peaks remain different: the maximum value of the peak increases with the frequency. This could occur if the charges at the barrier are coupled purely capacitively to the charges at the contacts, blocking DC currents but allowing current to flow at high frequencies. This remains speculative; further study is required. We thus choose 283 Hz as a compromise between distortion from measurement equipment and prominence of the conductance peak, the feature we focus our investigations on.\par

\section{Additional Drag Measurements}
\label{sec:AdditionalDrag}

\begin{figure}
	\centering
	\includegraphics[width=0.9\columnwidth]{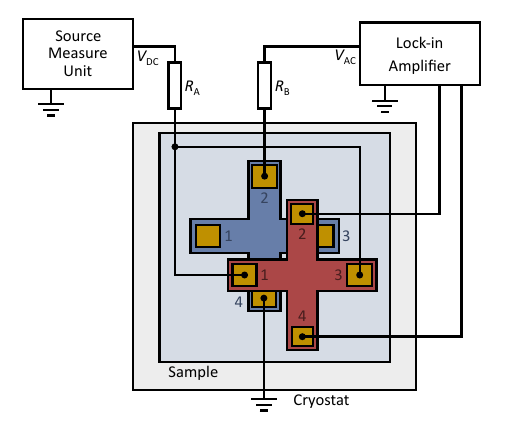}
	\caption{Setup for measuring drag voltages. In this example, the driving current is sent from contact 2 to 4 of the \textit{n}-layer. The drag voltage is picked up on the corresponding contacts, 2 and 4, of the \textit{p}-layer.}
	\label{fig:SM_DragMeasurementSetup}
\end{figure}

The setup used for drag measurements is shown in Fig.~\ref{fig:SM_DragMeasurementSetup}. As with the intralayer transport measurements (see section \ref{sec:IntralayerCircuit}), a current is sent through one of the layers using an AC excitation voltage provided by a lock-in and a large resistor ($R_{\text{B}}$). The drag voltage is then measured across the corresponding contacts of the other layer, again using the lock-in. The SMU is used, as usual, to apply a DC bias between the two layers. A resistor is added at the output of the SMU ($R_{\text{A}}$).\par

Variations on this measurement circuit were tested in order to confirm that the measured voltage conformed to the expectations for a Coulomb drag signal. In certain cases it did not. The remainder of this section describes those deviations, and how they informed the resulting final configuration that was used.\par

\begin{figure}
	\centering
	\includegraphics[width=0.9\columnwidth]{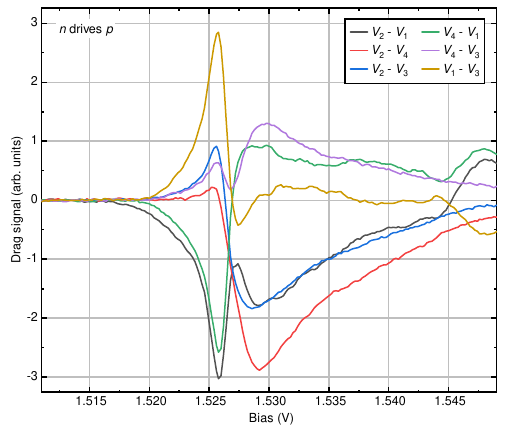}
	\caption{Drag measurements with different contact configurations.}
	\label{fig:SM_DragContactConfigs}
\end{figure}

Fig.~\ref{fig:SM_DragContactConfigs} displays voltages measured across different contacts of the dragged layer. Here there is no resistor at the SMU output, the contact at which the DC bias is applied is fixed (contact 1 of the \textit{p}-layer), and current is always driven from contact 4 to 2 of the \textit{n}-layer. The curves appear to represent two different signals: a sharp peak at lower biases and a broader peak at higher biases. The sharp peak is consistent with a voltage drop from contact 1 to 3. On the other hand, the broad peak is consistent with a voltage gain from contact 2 to 4. Thus, only the broad peak at higher biases appears consistent with a Coulomb drag signal from the underlying current in the \textit{n}-layer.\par

\begin{figure}
	\centering
	\includegraphics[width=0.9\columnwidth]{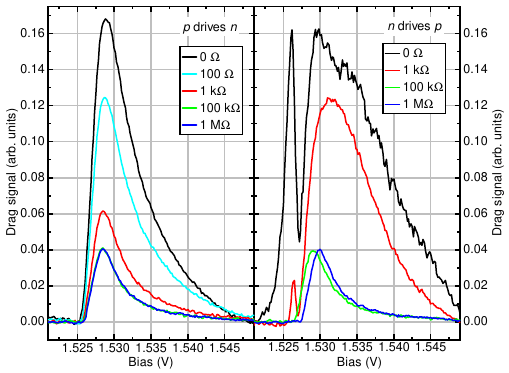}
	\caption{Drag measurements with different resistors at the SMU output. The left panel displays drag signals picked up in the \textit{n}-layer, while the right panel shows those measured in the \textit{p}-layer.}
	\label{fig:SM_DragSmuResistors}
\end{figure}

To probe the nature of the sharp peak, subsequent measurements were performed in which different resistors were added at the output of the SMU. The results are presented in Fig.~\ref{fig:SM_DragSmuResistors}. The SMU position was fixed (to contact 2), and the drag was always measured from contact 2 to 4 (with current driven through the same contacts in the opposing layer). For drag measured in the \textit{p}-layer (right panel), the sharp low-bias peak is suppressed with higher resistors at the SMU output. For drag measured in the \textit{n}-layer (left panel) the sharp peak is always absent. Note that the broad peak also initially deteriorates with increasing resistance, but remains constant for 100 k\textOmega\space and above. We therefore take this as the true Coulomb drag signal, and only present measurements with a 100 k\textOmega\space SMU resistor in the main text.\par

\begin{figure}
	\centering
	\includegraphics[width=0.9\columnwidth]{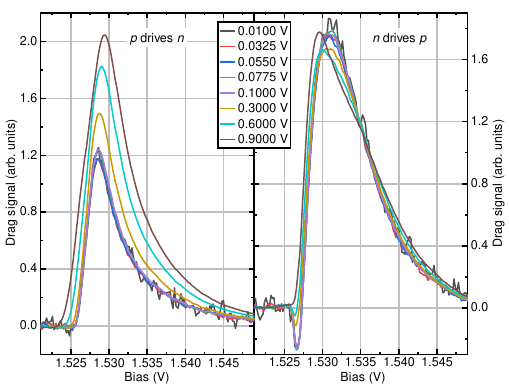}
	\caption{Drag measurements with different AC excitation voltage amplitudes. All measurements have been scaled by the voltage amplitude applied, so that in the linear regime all curves should coincide. This is the behavior seen for voltage amplitudes from 10 to 100 mV. The left panel displays drag signals picked up in the \textit{n}-layer, while the right panel shows those measured in the \textit{p}-layer. A 1 k\textOmega\ resistor was placed at the SMU output, which was subsequently attached to contacts 1 and 3, and the drag was measured across contacts 2 and 4 (with the current in the opposing layer driven also from contact 2 to 4).}
	\label{fig:SM_DragVacValues}
\end{figure}

Finally, we check that the drag signal is linear with the applied AC excitation voltage. Fig.~\ref{fig:SM_DragVacValues} shows that this is indeed the case for voltage amplitudes between 10 and 100 mV (with a 1 M\textOmega\ current-defining resistor). The increasing peak height at higher excitation voltages (i.e. AC currents) can be attributed to sample heating. At/near base temperature, dilution refrigerators have very small cooling power, and our highest AC currents (300 nA and up) exceed by far what is typically used in magnetotransport experiments in the millikelvin range (1-10 nA). The effect is more pronounced when driving currents through the \textit{p}-layer, as is expected, considering its higher resistivity.\par

\section{Drag Resistivity Correction}

In a typical Coulomb drag experiment, the drag resistance is taken as the voltage measured across the dragged 2D charge layer over the current sent through the driving 2D charge layer. The dragged layer is only connected to the voltage measurement device. In our case, the bulk-doped layers form conduction channels parallel to the 2D charge layers. Thus it is expected that the drag voltages we measure will be reduced compared to those in a typical experiment. We calculate a correction factor to account for this discrepancy.\par 

\begin{figure}
	\centering
	\includegraphics[width=0.9\columnwidth]{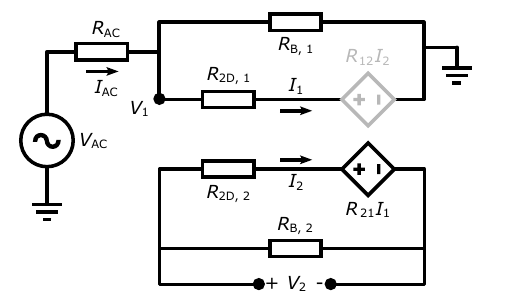}
	\caption{Circuit used to model the effect of parallel conduction channels ($R_\text{B, 1}$ and $R_\text{B, 2}$) on the measured drag resistance ($V_2 / I_\text{AC}$).}
	\label{fig:SM_DragCircuitModel}
\end{figure}

A circuit model as shown in Fig.~\ref{fig:SM_DragCircuitModel} is assumed for our system. The resistor $R_\text{AC}$ is chosen to be large enough that, together with excitation voltage $V_\text{AC}$, it sets the total current $I_\text{AC}$. The remainder of the circuit elements represent the device itself. $R_\text{B, i}$ and $R_\text{2D, i}$ represent the resistances of the bulk-doped layers and the 2D layers, respectively. The diamond-shaped voltage sources are the components modeling the drag effect and are dependent on the current through the opposite layer. $V_2$ is the measured drag signal. Without the bulk-doped layers, the drag resistance would be calculated as $V_2 / I_\text{AC}$.\par

This circuit was constructed according to the equations presented in \cite{RevModPhys.88.025003} (following \cite{pogrebinskii1977mutual}). These can be expressed with a matrix equation for the 2D layer currents and voltages, as shown in equation (\ref{eqn:DragMatrix}). We make the additional assumption that the drag effect is small enough that feedback on the first layer from the dragged layer current is negligible (the voltage source in gray in Fig.~\ref{fig:SM_DragCircuitModel} is ignored).\par

\begin{align}
\label{eqn:DragMatrix}
	\begin{bmatrix} V_1 \\ V_2 \end{bmatrix}
	&=
	\begin{bmatrix}
		R_{11} & R_{12} \\ R_{21} & R_{22}
	\end{bmatrix}
	\begin{bmatrix} I_1 \\ I_2 \end{bmatrix}
\end{align}

The remainder of the calculations follow standard circuit analysis. The resulting equation relating the `real' drag resistivity $\rho_\text{D}$ to the measured drag resistance $R_\text{D}$ is given by equation (\ref{eqn:DragCorrectionEqn}). $L$ is the layer length, and $W$ is the layer width. Thus, the measured drag resistance is multiplied by the correction factor $\alpha$ in order to calculate the drag resistivity.\par

\begin{align}
\label{eqn:DragCorrectionEqn}
	R_D &\coloneqq \frac{V_2}{I_{\text{AC}}} \\
	\nonumber &= \rho_D \frac{L}{W} \left( \frac{R_{\text{B, }1}}{R_{\text{2D, }1}+R_{\text{B, }1}} \right) \left( \frac{R_{\text{B, }2}}{R_{\text{2D, }2}+R_{\text{B, }2}} \right) \\
	\nonumber &= \frac{\rho_D}{\alpha}
\end{align}

Note that here again estimates for the 2D layer resistance values are required (see section \ref{sec:DensityAndMobility} for a description of how this is handled). In addition, there is a geometric factor (i.e. $L/W$) which must be accounted for. We use the spacing between the closest edges of the p-type contacts for $L$, and the arm widths of the cross for $W$, resulting in a value of 1.3 for $L/W$. As we cannot exclude that the EHB extends for a small distance underneath the Ohmic contacts, our value of 1.3 is a conservative estimate.\par